\documentstyle[spie]{article} 

\title {Adaptive Optics Spectroscopy: Preliminary Theoretical Results} 

\author{Jian Ge\supit{a}, Roger Angel\supit{a}, David Sandler\supit{a,b}, Chris Shelton\supit{a}, Don McCarthy\supit{a}, Jim Burge\supit{a} 
\skiplinehalf 
\supit{a}Steward   Observatory, The University of Arizona,\skipline 
Tucson, AZ \hspace{0.5em}85721 \hspace{0.5em}USA 
\skiplinehalf 
\supit{b}ThermoTrex Corporation, 10455 Pacific Center Court\skipline
San Diego, CA\hspace{0.5em}92121 \hspace{0.5em}USA
}

\authorinfo{Other author information: (Send correspondence to Jian Ge)\\J. G.: 
Email: jge@as.arizona.edu;  WWW: http://qso.as.arizona.edu/$^\sim$jge;
 Telephone: 520-621-6535; Fax: 520-621-1532}

\begin{document}
\maketitle 

\begin{abstract}
	   
Diffraction-limited spectroscopy with adaptive optics (AO) has several 
advantages over traditional seeing-limited spectroscopy. First, 
high resolution can be achieved without a large loss of light at the 
entrance slit of the spectrograph. Second, the small AO image width allows 
the cross-dispersed orders to be spaced closer together on the detector, 
allowing a large wavelength coverage. Third, AO spectrograph optics are slow
 and small, costing much less than for a traditional spectrograph. Fourth,
 small AO images  provide high spatial resolution. Fifth, scattered light is less problematic. And last,  the small entrance slit of the spectrograph 
can get rid of much of the sky background  to  obtain spectra of faint objects.

We have done theoretical calculations and simulations for infrared 
spectroscopy at the 
MMT 6.5 m with laser guide star AO, which provides almost full sky coverage.
The results show we can expect  40-60\% of the photons from a unresolved source 
within 0.2 arcsec diameter circle for J, H, K, L and M bands under typical atmospheric seeing condition at 2.2 micron
(r$_0$ = 1.0 m, t$_0$ = 21 ms, $\theta_0$ = 15 arcsec and d$_0$ = 25 m). 
Therefore,  the spectrograph  entrance slit size should match the 0.2 arcsec 
image to obtain high throughput. Higher resolution can be achieved by 
narrowing down the slit size to match the diffraction-limited image core size
 of about  0.1 arcsec in the infrared. However, the throughput will be 
correspondingly reduced by a factor of  two. Due to the limited atmospheric 
isoplanatic angle in the J, H and K bands, the 
encircled photon percentage within 0.2 arcsec diameter drops from 40-60\% when 
the object is at the laser pointing direction to 20-40\% when the object is 
about  30 arcsec away from the laser direction. Therefore, the useful field 
of view for AO multiple object spectroscopy is about 60 arcsec.  Further 
studies of  IR background (sky and thermal) and IR detector performance 
show that spectral resolution of R = 2,000 can take full advantage of AO 
images without much penalty due to the dark current of the IR detector and 
IR OH sky emission lines.

We have also studied natural guide star AO spectroscopy. Though sky coverage 
for this kind of spectroscopy at the MMT 6.5 m is very limited, a bright star
  provides much better performance than the laser guide star AO 
spectroscopy.  About 40-70\% photons are concentrated within 0.1 arcsec 
diameter  for guide stars  brighter than 13 magnitude. Therefore, higher 
resolution and high throughput can be obtained simultaneously, given a bright 
enough natural guide object.  The field-of-view for  multiobject spectroscopy 
using a natural guide star is similar to that for laser guiding.

\end{abstract}



\section{INTRODUCTION}

Adaptive optics (AO) promises revolutionary advances  in imaging  power for 
ground-based optical  and infrared astronomical telescopes
  by eliminating the wave-front  distortion caused by atmospheric  turbulence. 
The AO corrected images will be  nearly diffraction-limited,  which is about
a factor of ten times smaller than that limited by the atmospheric seeing for 
current 4 m class telescopes. For the largest of  the new generation of  
telescopes, the most dramatic gain  is possible, permitting an imaging 
performance of almost  two orders of magnitude  (100 times). 

Though adaptive optics has a big impact on improving ground-based telescope image
quality, it cannot provide  ideal diffraction-limited images in principle,
due  to the limited photon  flux  available from the reference source, finite  response time and subaperture size of  the  AO systems (Sandler et al. 1994).
The AO corrected images therefore consist of two components: a 
diffraction-limited core and a broad seeing-limited halo (Beckers  1993), 
which make the  design of AO  instruments different from   that of seeing
limited instruments. 

The  much sharpened  AO images have two main applications
in astronomy research; imaging and spectroscopy. The two  are closely related 
but not  the same. The main focus of  direct high resolution imaging is  to 
sharpen  the diffraction limited image core, to maintain stable uniform
 point spread function (PSF) in both spatial and temporal domains.  On the other
hand, the most  concern of the  AO spectroscopy is the flux concentration. The 
different demands for these two different  applications determine 
different instrument design parameters.  

The application of adaptive  optics in astronomy is still in its early phase 
and  the design of  AO optimized instruments, especially the  spectrographs,
 is a new territory being opened for exploration. In seeing 
limited domain, the best resolution of a spectrograph   is coupled  to 
the  telescope diameter, the larger the aperture size, the lower the spectral 
resolution for the normal available grating size. This coupling 
 limited the best spectral resolution of traditional spectrographs 
to R $\sim$ 50,000 for 4 m class telescopes (Vogt \& Schroeder 1987). In order 
to obtain higher resolution, all kind of tricks such as image slicers, pupil 
slicers, grating mosaics etc have been applied  and 
resulted in very large and expensive
spectrographs at the Nasmyth  or Coude focuses (e.g. 
Diego  et al. 1995; Vogt et al. 1994; Tull et al. 1994). However, in the 
AO diffraction limited domain,  because the AO corrected 
image size (i.e. diffraction limited core size) decreases proportionally 
with telescope aperture size, the coupling of  spectrograph size with 
telescope aperture  is removed. Very high resolution 
spectrographs  can be made from normal size gratings. As the results, the 
next generation AO optimized high resolution spectrographs  will have smaller
overall scale,   higher efficiency  and also cost much less. 
 As  the first demonstration of next generation AO spectrographs, we have
built a prototype AO cross-dispersed echelle spectrograph with  a
125x250 mm$^2$ R2 echelle grating at Steward Observatory and tested  at Starfire
Optical Range 1.5 m AO telescope.  The spectrograph can provide
 spectral resolution up to  
R = 700,000  (Ge et al. 1996).   Because of the much smaller image size, 
 a large  amount of cross
dispersed orders can be packed and recorded on the detector, and thus a factor 
of 100 times larger  wavelength
coverage over similar resolution traditional spectrographs was
achieved (e.g. Diego  et al. 1995; Lambert  et al. 1990; Ge et al. 1996). And  
because  of the much smaller image size, AO spectrographs can record 
astronomical phenomena of much smaller scale structure  (Bacon et al. 1995).
Further,  smaller
entrance slit  used in the AO  spectrographs can  help  to  block most 
sky background, especially in the IR where the sky
background is about 100 times brighter than in the visible, so much fainter objects can be observed.

In this paper, we will first set out the types of error  that arise in the 
adaptive optics systems and how  they together affect overall performance of 
 AO spectroscopy. Then we will use the MMT 6.5 m  AO system under 
construction as an example to  introduce the results from the semi-empirical analytical  calculations and  direct Monte Carlo  simulations 
and  relate these computational results to the 
design of AO spectrographs. 


\section{Theoretical Considerations}

Adaptive optics residual error contributions can be divided into tilt and 
higher order  error terms (Parenti 1992; Sandler et al. 1994). AO systems 
 often apply corrections to these aberrations  separately. The low  order 
tilt errors are compensated by  the tilt-control element and the high 
order errors  are  corrected by the  deformable mirror. 
 Therefore the overall image quality can be described as 
\begin{equation}
S_{tot} = S_{tilt} S_{HO}
\end{equation}
where $S_{tot}$ is the total system  Strehl ratio (SR), $S_{tilt}$ is the SR for
the tilt error correction only and $S_{HO}$ is the SR  for the high-order   
aberration  correction only.  

\subsection {Tilt Errors}

The tilt, caused by the effects of atmospheric jitter,  has a longer 
correlation time due to its large scale. Three major sources
contribute  to the residual tilt errors: anisoplanatism ($\sigma_{iso}$),
 centroid  uncertainty ($\sigma_{cent}$)
and temporal decorrelation ($\sigma_{temp}$) (Sandler et al. 1994). 
The total one-axis   mean square tilt error can be written as
\begin{equation}
\sigma_{tilt}^2  = \sigma_{iso}^2 + \sigma_{cent}^2 + \sigma_{temp}^2.
\end{equation}

The tilt  anisoplanatism  is caused  by the different atmospheric wedge 
traced by the light from the tilt  sensing  field  star and from the science  
object. It can be described as the average of tilt errors in the longitudinal
and  the lateral  directions,
\begin{equation}
\sigma_{iso}^2 = \frac{(\sigma_{iso}^x)^2 + (\sigma_{iso}^y)^2}{2},
\end{equation}
where  the longitudinal tilt anisoplanatism,
\begin{equation}
(\sigma_{iso}^x)^2  = [0.0472(\frac{\theta}{\theta_0})^2(\frac{D}{r_0})^{-1/3} - 0.0107(\frac{\theta}{\theta_0})^4(\frac{D}{r_0})^{-7/3}](\frac{\lambda}{D})^2,
\end{equation}
and  the lateral tilt anisoplanatism,
\begin{equation}
(\sigma_{iso}^y)^2  = [0.0157(\frac{\theta}{\theta_0})^2(\frac{D}{r_0})^{-1/3} - 0.00214(\frac{\theta}{\theta_0})^4(\frac{D}{r_0})^{-7/3}](\frac{\lambda}{D})^2,
\end{equation}
where $r_0$ is the atmospheric turbulence coherence length, and  $\theta_0$
is the size of the isoplanatic angle, D  is the telescope aperture  size  and 
$\lambda$ is  the  tilt sensing wavelength (see Sandler et al. 1994 for 
details).

The temporal  tilt decorrelation is caused  by the delay between measuring and
compensating for the atmospheric wedge. The average mean-square temporal
tilt error can be expressed  as
\begin{equation}
\sigma_{temp}^2 = 0.031 (\frac{T}{t_0})^2(\frac{r_0}{D})^{1/3}(\frac{\lambda}{D})^2,
\end{equation}
where $t_0$ is the atmospheric coherence  time  and $T$ is the delay time  
between sensing  and correcting.

Due to  the limited  photon flux available from the tilt field star  and 
detector intrinsic noise, the exact  centroid of the tilt guide star image
cannot be precisely measured. The resulting error is  called the tilt 
centroiding error, which can be  written as  
\begin{equation}
\sigma_{cent}^2  = \frac{\alpha^2 w^2}{N} (1+\frac{4n^2}{N})(\frac{\lambda}{D})^2,
\end{equation}
where the parameter $\alpha$ depends on $D/r_0$, $\alpha$ = 0.76 - 1.11 for  
$D/r_0$ = 7-10, $N$ is the total detected photon  number, $n$ is the detector
readout  noise, $w$ is the FWHM of the image in units of the diffraction-limited
image width $\lambda/D$. 

\subsection{High Order Errors}

The contributions to high order  correction errors are fitting ($\sigma_{fit}$),
reconstruction ($\sigma_{rec}$), temporal-decorrelation ($\sigma_{time}$),
  high-order anisoplanatism ($\sigma_{hoiso}$)  and focus anisoplanatism 
($\sigma_{cone}$) if laser guide  star is used. The total mean-square 
high-order correction error  is 
\begin{equation}
\sigma_{HO}^2 = \sigma_{fit}^2 + \sigma_{hoiso}^2 + \sigma_{rec}^2 + \sigma_{time}^2 + \sigma_{cone}^2.
\end{equation}

Because of its finite number of  actuators, the spatial frequency 
components with  scales smaller than a  subaperture size cannot
be corrected by the adaptive deformable mirror.  The resulting  wave-front 
error is called fitting error, which is further written as
\begin{equation}
\sigma_{fit}^2 =  c(\frac{d}{r_0})^{5/3},
\end{equation}
where $c \approx 0.29$, depends  on the actuator arrangement  and $d$ is the 
subaperture size  (Greenwood \& Parenti, 1994; Ellerbroek, 1991; Rigaut 1994).

High-order anisoplanatism arises from the angular  separation  $\theta$ 
between the reference source being sensed and  corrected and the object being 
imaged. A good approximation for this high order anisoplanatism  is
\begin{equation}
\sigma_{hoiso}^2 = \sigma_\phi^2 -ln[1+0.9736E+0.5133E^2+0.2009E^3+0.0697E^4+0.02744E^5],
\end{equation}
where 
\begin{equation} 
E = 6.88\frac{\mu_2}{\mu_0}(\frac{\theta}{D})^2(\frac{D}{r_0})^{5/3},
~~~~\sigma_\phi^2  = (\frac{\theta}{\theta_0})^{5/3}, ~~~~\mu_n =sec^{n+1}(\zeta) \int^\infty_0 h^n C_n^2(h) dh,  
\end{equation}
$C_n^2(h)$ is  the index-of-refraction structure constant at altitude  h,
 $\zeta$ is the
zenith angle (Greenwood \&  Parenti  1994; Sasiela 1993). This approximation
gives  accurate   Strehl  ratio values down to values of 0.2. 

The high order temporal decorrelation  error is  caused  by the change of
 the atmospheric  wedges over  each subaperture over the wavefront sensor delay time, $\Delta t$.  The error is approximately  
\begin{equation}
\sigma_{time}^2  = (\frac{\Delta t}{t_0})^{5/3}.
\end{equation}

The reconstruction  error has  the same general form as  the centroiding error
for tilt correction because the errors  introduced  by centroiding uncertainty
due to the photon noise and detector noise over each subaperture result in 
 a rms wavefront error in the final reconstructed wave front. 
It  can be written as  
\begin{equation}
\sigma_{rec}^2 = \frac{4\pi^2  G \alpha^2 w^2}{N}(1+\frac{4n^2}{N})(\frac{\lambda_0}{\lambda})^2,
\end{equation}
where G $\approx$ 0.5, which depends on  the geometry  of the reconstructor,
$\lambda_0$ is   the sensing wavelength  and $\lambda$ is the science  
wavelength.

Focus anisoplanatism is the effect where light  from a laser  beacon (finite 
height) does  not  sample  the same turbulence  as the light from the 
observed target  does (infinite height). The cone error is expressed as
\begin{equation}
\sigma_{cone}^2 =  (\frac{D}{d_0})^{5/3},
\end{equation}
where $d_0$ is a  length-like quantity defined by Fried  (1994).
 
\subsection{The AO Corrected Images for Spectroscopy}

Because of the existence of the low ($\sigma_{tilt}$) 
and high order  residual errors ($\sigma_{HO}$) after 
the AO correction, the AO corrected images are not perfectly  diffraction limited.  They generally consist of two components,
a diffraction limited core  and a broad uncorrected $``$seeing" halo. 
The diffraction limited core appears after the low   spatial frequency 
wavefront errors (such as tilt error) are reduced.
The halo components  decrease once the higher spatial 
frequency errors  are reduced. Most previous astronomical AO   systems are
low order   systems. Though they can provide  diffraction-limited image core
for high resolution imaging, the energy concentration within small angular
diameter is relatively low, therefore these  systems are  not ideal for 
AO spectroscopy. However,  the existing  high order  AO  systems at the
SOR 1.5 m telescope and Mt.  Wilson 2.5 m telescope have  the power to
 concentrate about 50\% photons within  central  0.2$''$. They are certainly
suitable for AO  spectroscopy  (Ge et al. 1996). Moreover,
 some new  AO  systems being  designed and built at large telescopes
 such  as the MMT 6.5 m will provide high order compensation which 
 can largely improve the photon  flux concentration
ability in the IR for  high throughput AO spectroscopy.

\section{Theoretical Study Results}

A number of  useful methods have been applied to  study the AO performance
  including direct
Monte Carlo computer simulation  of the whole AO  system (Sandler et al. 1994),
 semi-empirical analysis (Parenti 1992; 
Ridgway 1994) and the evaluations  in terms of residual mean-square phase  distortion   and  the  associated  optical transfer function (OTF)   (Ellerbroek  1994;  Ellerbroek et al. 1994).   In this paper, we will apply the first  
two methods to  analyze the  performance of the AO  system under
 construction for  the MMT 6.5 m, and to relate them to the AO spectrograph 
design.

\subsection{Semi-empirical Calculations}

Previous study by Parenti (1992) demonstrates  that
in a long-exposure image corrected  by an AO system 
the randomly fluctuating sidelobes appearing in 
a short-exposure image smooth out to form  a broad  quasi-Gaussian shape 
background skirt, which is 
determined by uncorrected beam motion. The diffraction-limited primary
lobe shown in the short integration forms another sharp quasi-Gaussian profile  
overlapped on the broad background profile. 
The PSF  of the resulting image
in  a long exposure   time is the sum of these two 
quasi-Gaussion  functions. Therefore, important parameters  to describe  the AO 
performance such as   Strehl  ratio, encircled energy and resolution can be
expressed  by analytical formulae which make the system performance 
study much easier than through other methods. For the most intermediate cases,
 the results from this simple 
method agree reasonably well with much complicated computer-simulation  results 
as shown in the next  subsection and agree with the real time  observation 
results as  well. However, for the very poor and very good correction 
cases, the results from this method represent the expected asymptotic behavior.

 In  this approach,    the width of the  AO corrected diffraction limited 
core  is expressed as 
\begin{equation}
W_c = \sqrt{(1.22\frac{\lambda}{D})^2 + (2.7\sigma_{tilt})^2},
\end{equation}
and the width of the uncorrected halo can be written as 
\begin{equation}
W_h = 1.22 \frac{\lambda}{r_0}.
\end{equation}
The central intensity of the core component is 
\begin{equation}
I_c = \frac{exp(-\sigma_{HO}^2)}{1  + 4.94(\frac{D}{\lambda})^2\sigma_{tilt}^2},
\end{equation}
and the peak  intensity of the halo component is
\begin{equation}
I_h = \frac{1- exp(-\sigma_{HO}^2)}{1  + (\frac{D}{r_0})^2}.
\end{equation}
Therefore, the PSF can be approximately expressed  by (Ridgway 1994)
\begin{equation}
I(\alpha)=  I_c  exp(-\frac{4ln2}{W_c^2}\alpha^2) + I_h exp(-\frac{4ln2}{W_h^2}\alpha^2),
\end{equation}
where  we have assumed  Gaussion shapes for both components,  $\alpha$ is the 
angle from the image center. With this  definitions, the Strehl ratio  is the
value in Eq. 19  when $\alpha$ = 0, i.e.
\begin{equation}
SR = I(0) = I_c + I_h.
\end{equation}
The encircled  energy within $\beta$ angle is  the integral  of $I(\alpha$), or
\begin{equation}
E(\beta) = \int^\beta_0  I(\alpha)  2\pi \alpha d\alpha = \frac{1}{W_c^2I_c + W_h^2 I_h}[W_c^2I_c(1-exp(-\frac{4ln2 \beta^2}{W_c^2}))+W_h^2I_h(1-exp(-\frac{4ln2 \beta^2}{W_h^2}))].
\end{equation}

 As an example, Table 1 shows direct comparisons between the theoretical 
predictions from the semi-empirical analysis and observation results 
 from the Mt. Wilson 2.5 m telescope
 AO system. The atmospheric parameters at 2.2 $\mu$m, r$_0$ =  1.8  m 
or 0.4$''$ seeing, wind velocity 21 m s$^{-1}$ or t$_0$ = 15 ms  were 
applied in order to match the theoretical predictions
 with the observation values in the R band.  
Subaperture size of 0.156 m, wavefront sensor delay
 time of  5.0 ms and tilt correcting delay time of 5.0 ms  were 
also used in the calculations. The observed star is SAO 140094 with V = 6.6 
mag.   The resulting predicts from the first  order theoretical analysis  
also reasonably  match with  the observation  values in the other
 wavelength regions such as I, V and  B bands. 
 
\begin{table} [h]   
\caption{Comparison between Theoretical and Observational Results for the Mt. Wilson 2.5 m Telescope NGS AO System.} 
\begin{center}       
\begin{tabular}{|c|c|c|c|c|c|c|} 
\hline  
\rule{0pt}{2.5ex} 
  Wavelength & SR(the)& SR(ob) &0.1$''$EE(the) &0.1$''$EE(ob)&0.2$''$EE(the)&0.2$''$EE(ob)\\[0.2ex]
\hline  
I(0.9$\mu$m)&0.34&0.36&22\%&22\%&46\%&42\%\\
R(0.7$\mu$m)&0.18&0.17&17\%&17\%&33\%&35\%\\
V(0.55$\mu$m)&0.07&0.09&10\%&17\%&23\%&31\%\\
B(0.44$\mu$m)&0.02&0.03&6\%&8\%&18\%&22\%\\[0.2ex]
\hline  
\end{tabular}
\end{center}
\end{table}

In the following, we  applied the semi-empirical 
formulae to explore the on-axis and  off-axis performance of the MMT 6.5 m  
LGS and NGS  AO systems.
In the calculations, the laser beacon was assumed to point in the
direction of the science object for high order correction and a H = 18 mag. 
field star 30$''$ away from  the science object was used for the global  tilt 
correction. The LGS therefore can provide image corrections  for almost
 anywhere on the sky (Sandler et al. 1994). We further assumed that the wavefront 
sensor  sensing and compensating time   is 2 ms and global tilt  time delay
 is  15  ms.  The average values of the atmospheric parameters,  
r$_0$ = 1.0  m, t$_0$ = 21.2 ms,  d$_0$  = 25 m  and $\theta_0$ = 15.4$''$ at 2.2 $\mu$m were adopted. 
The laser is a  4 watts  sodium laser, which is applied to the mesosphere
sodium layer to  form  a R  =  9.5 mag. artificial sodium star. 
This is certainly achievable  with the cw dye laser we are using  at the MMT.
For   instance,  recent  simultaneous measurements of  the
sodium laser  guide star return and mesospheric sodium column density at the
MMT and CFA 60 inch telescopes  on Mt. Hopkins  show that a  R = 10
mag. laser guide star formed from  1  watt  projected sodium laser   power 
on the sky  when the sodium  column  density is about  the  annual mean value of
3.6$\times 10^9$ cm$^{-2}$ (Ge  et al. 1997). 
Table 2 shows the calculation results about the LGS AO system 
from the above empirical formulae. 
  
\begin{table} [h]   
\caption{The  Calculation Results for the MMT 6.5 LGS AO  System Performance  
under the Typical Seeing Condition on Mt. Hopkins.} 
\begin{center}       
\begin{tabular}{|c|c|c|c|} 
\hline  
\rule{0pt}{2.5ex} 
  Bands & Strehl Ratio & 50\% Encircled Energy  & 80\% Encircled Energy\\[0.2ex]
\hline  
   	J(1.25$\mu$m) & 0.29 & 0.32$''$& 0.75$''$ \\
	H(1.65$\mu$m) & 0.40& 0.12$''$ & 0.59$''$ \\
	K(2.2$\mu$m) & 0.48 & 0.12$''$ & 0.36$''$ \\
	L(3.6$\mu$m) & 0.56 & 0.16$''$ & 0.26$''$ \\
 	M(4.8$\mu$m) & 0.58 & 0.20$''$ & 0.32$''$ \\[0.2ex]
\hline  
\end{tabular}
\end{center}
\end{table}

Figure 1 shows the results of LGS AO corrected image angular diameter
 versus wavelength in 
the near IR  (1 - 5 $\mu$m).  The
photon flux  concentration strongly depends on wavelength. At the J band, 
uncorrected high order wavefront errors especially the focus  anisoplanatism
caused by the single laser beacon are significant, uncorrected $``$seeing" 
halo dominates  the whole image,  only about 30-40\% photons
are  concentrated  within 0.2$''$ central image area. In the H and K, the AO 
corrected diffraction-limited component begin to be dominant, 50-60\% photons are
within the same 0.2$''$ angular diameter. As the wavelength moves to 
 the  L and M bands, the AO corrected  images are close
to the diffraction-limit. However, the diffraction-limited core size  
is also getting larger at these longer wavelengths. Therefore 
there  are about 40-50\% photons  concentrated within 0.2$''$ aperture. 

	\begin{figure}	
	\vspace{9.5cm}	
\includegraphics{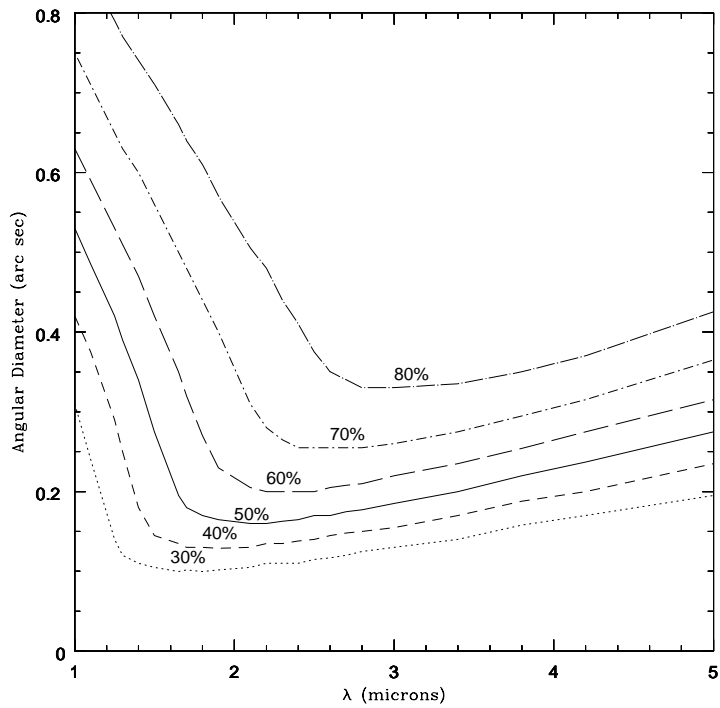}
	\caption
	{The image angular diameter vs wavelength at the near
IR for the MMT LGS AO system  under the average seeing conditions. Different
fractional  encircled  energy lines are drawn and  marked. } 	
	\end{figure} 

	\begin{figure}	
	\vspace{9.5cm}	
\includegraphics{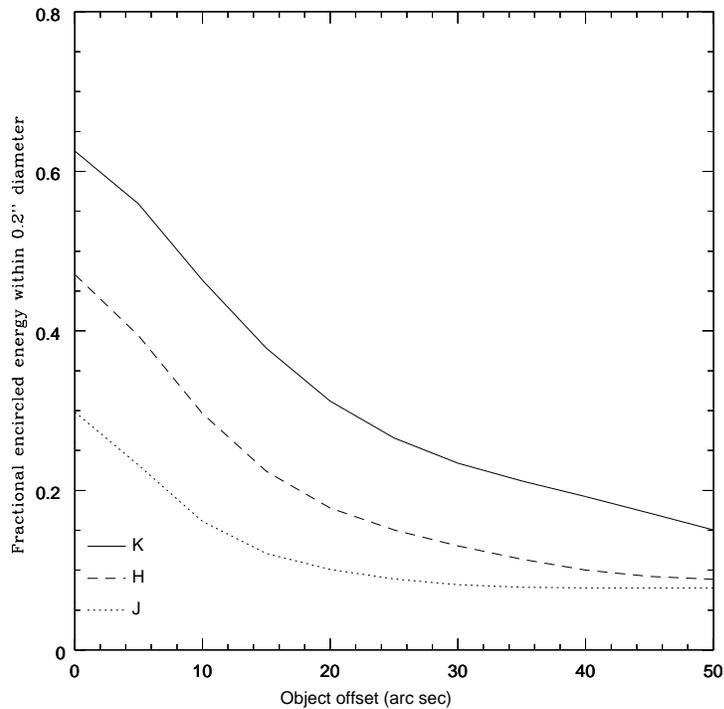}
	\caption[] 
	{Fractional encircled energy within 0.2$''$ diameter at the J, H and K
bands vs  the angle between the  laser  beacon and science object 
for the MMT LGS AO system  under the average seeing condition. } 	
	\end{figure} 

Figure 2 shows the off-axis performance from the LGS AO system, which 
illustrates the relatively narrow  corrected field-of-view.  However, the
corrected FOV strongly depends on the science wavelength. The  useful 
FOV for  J, H and K bands is generally less than 60$''\times 60''$, though
the corrected FOVs in the L and  M bands are expected to be larger.

Though the LGS AO system  dramatically  increases the sky coverage for  
science observations, a relatively wide  slit size of 0.2$''$  is required
to maintain high throughput for the IR spectroscopy. In order to  obtain
high spectral resolution such as R $\sim$ 100,000, the IR spectrograph 
system turns out  to be very large, which begin to push   limits on
  designs of the IR cryogenic and   mechanical systems. On the other hand,
 the NGS 
AO system is expected  to perform much better in the short IR wavelength
 than the LGS AO  because  there is no focus anisoplanatism contribution to the 
residual high order wavefront aberrations.  
Figure 3 shows  the calculation results of the MMT  NGS AO performance  under
the average seeing conditions. As expected, 
 the fractional energy concentration  is much higher than that from the LGS AO system. For instance, the  fractional encircled 
energy is about 20-40\% within 0.1$''$ angular diameter for a NGS 
with V  $\sim$  14 magnitude, and  40-70\% for a  NGS guide  
star brighter than 13 magnitude. Therefore, in the very limited observation 
case when there is a bright NGS in the  field, especially when 
the science object
is the NGS itself, IR spectroscopy can be pursued with a $\sim$ 0.1$''$ entrance
slit, a factor of two higher spectral  resolution can therefore be achieved.

	\begin{figure}	
	\vspace{9.5cm}	
\includegraphics{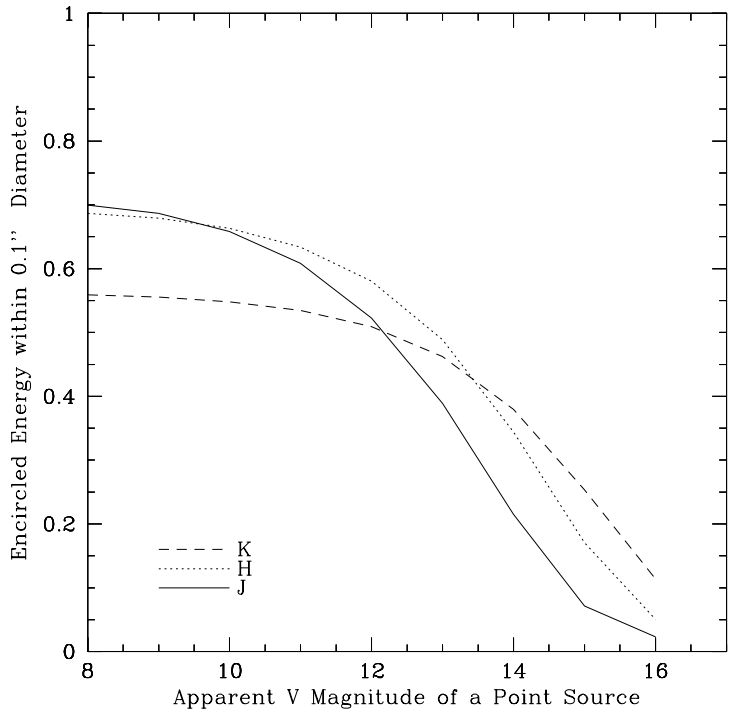} 
	\caption[] 
	{Fractional encircled energy vs apparent V magnitude  of a
point  source with the MMT 6.5 m NGS AO  system under  the average seeing  condition.  } 	
	\end{figure}

\subsection{Monte  Carlo Simulations}

This method has been used  to  study the LGS AO performance on 8-m  telescopes
(Sandler et al. 1994).  The basic procedure is  to generate a perfect 
plane wavefront and then let it pass through  10  to 20
 Kolmogorov  phase  screens spaced between 3 to 20 km. The  global tilt in the 
simulated distorted  wavefront from a H = 18 mag. field star was   sensed 
in the wavelength of 1.25-2.2 $\mu$m and the measured wavefront slopes across 
the 6.5 m aperture were used to   remove the wavefront tilt from the science
object, which is 30$''$ away. Then 
the high order aberrations sensed by a simulated  back-scattered 
laser beacon from the sodium  layer (90 km altitude) in the direction 
of the science  object  were measured and  applied  to provide  high order correction for the  wavefront of the science object.  The deformable mirror
used for the  high order correction  has 13x13 actuators  with the subaperture 
size of 0.5 m.  Other parameters used in  this simulation approach are  the 
same as used in the above semi-empirical analysis.   The output 
PSFs were used to measured the resulting Strehl ratio, fractional 
encircled energy, resolution etc. 

In the simulations, we used  the MK atmospheric model described in the paper by 
 Sandler et al. (1994), corresponding to the atmospheric parameters of 
r$_0$ = 1.0 m, t$_0$ = 21 ms, d$_0$ = 25 m and $\theta_0$ =  15.4$''$ at
 2.2 $\mu$m, typical values on Mt. Hopkins.  Figure  4 shows example
  image profiles in the K band for  the
diffraction-limited  and  the LGS AO  corrected cases from the
 simulations. The Strehl ratio
for the  AO corrected   image is 0.55. Compared to the ideal diffraction-limited
image profile, the AO sharpened image shows that more photons spread out in the 
wing of the image profile. The whole  profile indeed consists  of   a sharp
Gaussian profile  with the FWHM similar to the diffraction-limited core width
 and  a much broader Gaussian profile (Parenti 1992).   
The broad seeing-like  halo profile is caused by the residual 
 high spatial frequency aberrations mainly contributed  by the 
LGS focus anisoplanatism.

\begin{figure}	
\vspace{9.5cm}	
\includegraphics{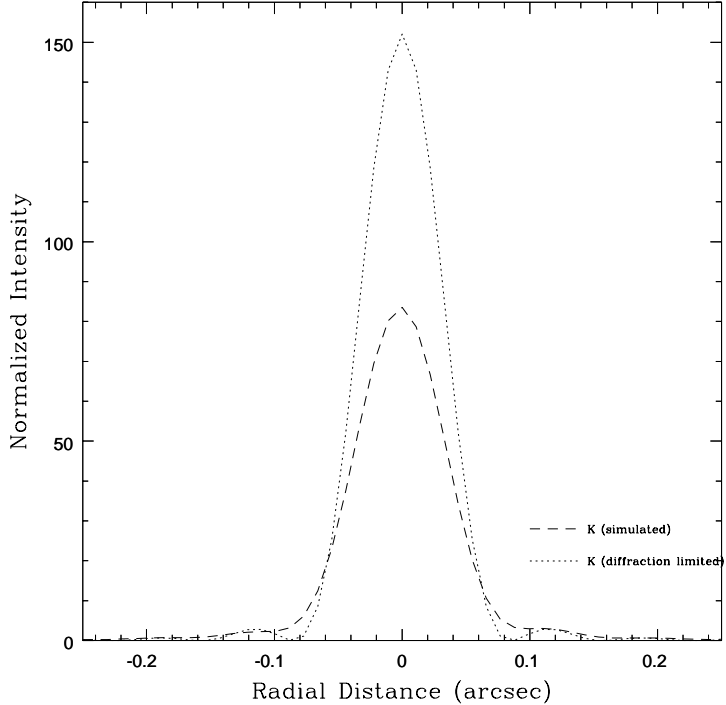} 
\caption[]
{Simulated K band  diffraction-limited and
laser guide star AO  corrected stellar image profiles from the  MMT 6.5 m.} 	
\end{figure} 

 Figure 5 shows the fractional encircled energy versus angular diameter from the
simulations. MMT 6.5 m LGS AO 
system can concentrate about 50\% photon within 0.2$''$ diameter
for wavelength longer than 1.4 $\mu$m. Once  the  correction  goes to
the shorter wavelength, the laser cone effect and fitting error  become
significant,  and  less photons are
within the diffraction  limited core. Figure 5 further illustrates that  the
rest $\sim$ 30\% uncorrected photons  from  the MMT 6.5 m single beacon LGS AO 
system   are widely distributed  over a  large area.  This new  result
 has important meaning  for the IR spectrograph design. 
	\begin{figure}	
	\vspace{9.5cm}	
\includegraphics{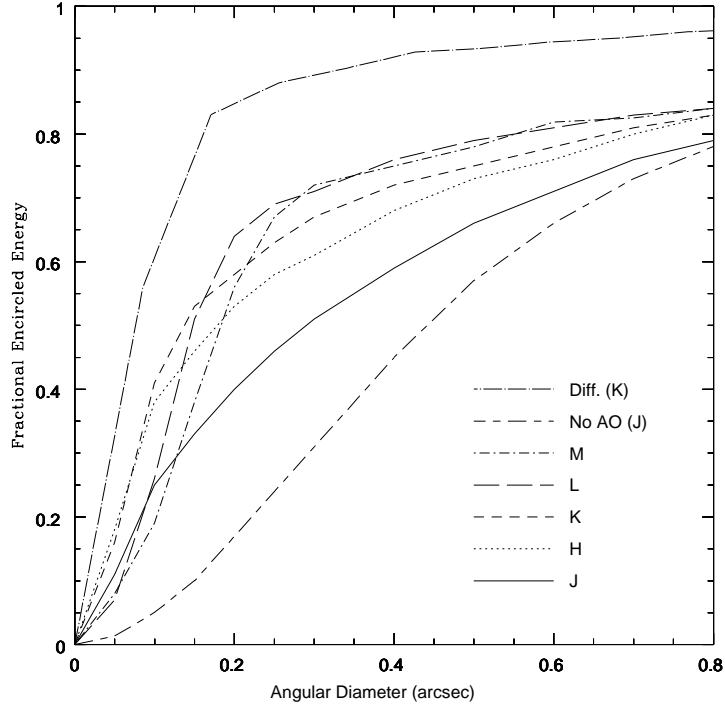} 
	\caption[] 
	{Fractional encircled energy vs angular diameter  for the MMT LGS AO system
under average seeing condition. } 	
	\end{figure} 

Table 3 summarizes  the  simulation results in the J, H,  K, L and  M  bands. 
Compared to the semi-empirical calculation results  shown in the Table 2, 
the simulations show similar results  on the  50\% 
encircled energy  in the whole near IR wavelengths, but gives lower  Strehl 
ratios for the J and H bands, anf higher SRs for the K, L and M bands, and
also  predicts different  aperture sizes for the 80\% encircled energy.

\begin{table} [h]   
\caption{The simulation results for the MMT 6.5 m LGS AO  system performance  
under typical seeing condition on Mt. Hopkins.} 
\begin{center}       
\begin{tabular}{|c|c|c|c|} 
\hline  
\rule{0pt}{2.5ex} 
  Bands & Strehl Ratio & 50\% Encircled Energy  & 80\% Encircled Energy\\[0.2ex]
\hline  
   	J & 0.16 & 0.28$''$& 0.80$''$ \\
	H & 0.36 & 0.19$''$ & 0.73$''$ \\
	K & 0.55 & 0.13$''$ & 0.65$''$ \\
	L & 0.79& 0.15$''$ & 0.50$''$ \\
 	M & 0.88 & 0.18$''$ & 0.50$''$ \\
	J(no  AO) & 0.02  & 0.45$''$  & 0.90$''$\\[0.2ex]
\hline  
\end{tabular}
\end{center}
\end{table}

\section{Design of AO Spectrographs}

The preliminary results  from the computer simulations and  theoretical 
calculations demonstrate that astronomical adaptive  optics  with laser or
natural guide stars cannot achieve ideal diffraction-limited  images even 
at the longer IR wavelength  region such as L and M bands. The shapes of the 
resulting AO corrected image profiles depend on  how well the 
low order and high order wavefront aberrations have been corrected, 
which can be easily affected by the instantaneous atmospheric conditions,
guide star magnitudes  and nature (LGS or  NGS), and also AO correcting parameters. All these make  the design of AO instruments, especially 
spectrographs, in some respects   more complicated than for the seeing limit. 

As shown above, in order to  provide full sky coverage of the AO observations, 
laser guide star is required. 
The extra high order aberration from the LGS,
focus anisoplanatism, is the dominant one 
in  the whole  IR wavelength range and will make a large fraction of  photons  spread out in the broad wing
 portion of  the resulting corrected PSF.  About  40-60\% photons are within 
$\sim$ 0.2 $''$ angular diameter, or 2-3 diffraction-limited core size. 
  Therefore, the optimal design of the AO spectrographs is to match the
 entrance slit width to the 2-3 diffraction-limited core size  to provide 
high throughput. A smaller width slit could be applied to obtain high spectral 
resolution, but the throughput dramatically  decreases as  the slit narrows
 down. On the other hand, slightly 
larger than 0.2$''$ slit size cannot improve  much more throughput due to  the 
wide distribution of the uncorrected photons.  Moreover, larger 
slit   size will bring in more  sky  and thermal background into the 
spectrograph system  and decrease the detectability and spectral 
resolution. 

Because of the low  and high  order anisoplanatism, the  AO corrected  
field-of-view is limited  to about 1 arc min, and image quality degrades 
dramatically at the edge of the field. However, the relatively small
 AO  corrected FOVs could be still  very  useful  for  the multiple object
 IR AO spectroscopy because the largest  size of the present available
 IR detector such  as   the 1kx1k InSb array naturally matches the AO 
corrected FOVs (Fowler et al. 1996). 

Though sky coverage by bright natural  guide stars with V $\le$ 14  mag. is 
very small, the image quality corrected  by the NGS is much better than that
by the LGS.  For this special case, about 40\% photons are  within 0.1$''$ 
diameter.  So much narrower slits could be  used to maintain the high 
throughput, but increase the spectral resolution of about  a factor of  two.

Together, the design of the IR AO spectrographs should at least coordinate 
the needs of the  0.1$''$, 0.2$''$ entrance aperture  sizes for the LGS and
NGS, respectively. Further, in about
10\% best seeing conditions, the AO corrected images will be very close to the 
ideal  diffraction  limited, much narrower slit  width  matching the 
diffraction limited  core size, e.g. 0.06$''$ in the H band for the MMT 6.5 m,
 could be used. Including this option in the  spectrograph design could 
significantly increase the spectral resolution.

As we mentioned before, another big  potential advantage with the 
IR AO spectroscopy  is that  much smaller spectrograph entrance aperture 
size can help block most sky background  which is always associated with the 
seeing-limited IR spectroscopy. A factor of 5-10 times fainter limit can be 
reached with the AO spectroscopy if sky background is the dominant one. However,
there is always   dark current associated with the IR detector, the lowest is
about $\sim$ 0.1 e s$^{-1}$ (Fowler et al. 1996). If  the 
sky background  gets too much dispersed by the spectrograph gratings, then the
detector noise, especially dark current, will be the dominant background 
noise.  The extra advantage of smaller slit width to the IR spectroscopy
 will eventually disappear. Detailed studies of IR background including 
sky OH emission lines and airglow emission  and  thermal emission show 
 that R $\sim$ 2,000 AO  spectroscopy can  take full  advantage  of
the  narrower slit width  (Ge et al. 1997 in preparation).

Another potential concern for the IR AO spectroscopy is the 
 atmospheric differential dispersion, which has been neglected in the seeing limited IR spectroscopy.  Figure  6  shows the atmospheric differential 
dispersion for different zenith  angles  at  different  wavelengths  above  Mt.
Hopkins.  The  
dispersed  images from the J and  K bands could be separated as large as $\sim$
0.2 $''$ if the observations are made at high airmasses.  Therefore,
 the slit loss caused  by this dispersion effect could be serious if
the spectroscopy covered a  large wavelength region. An atmospheric  dispersion 
corrector is needed to correct this effect, or 
instrument rotator is required to allow all the dispersed images  align
on the slit.

\begin{figure}	
\vspace{9.5cm}	
\includegraphics{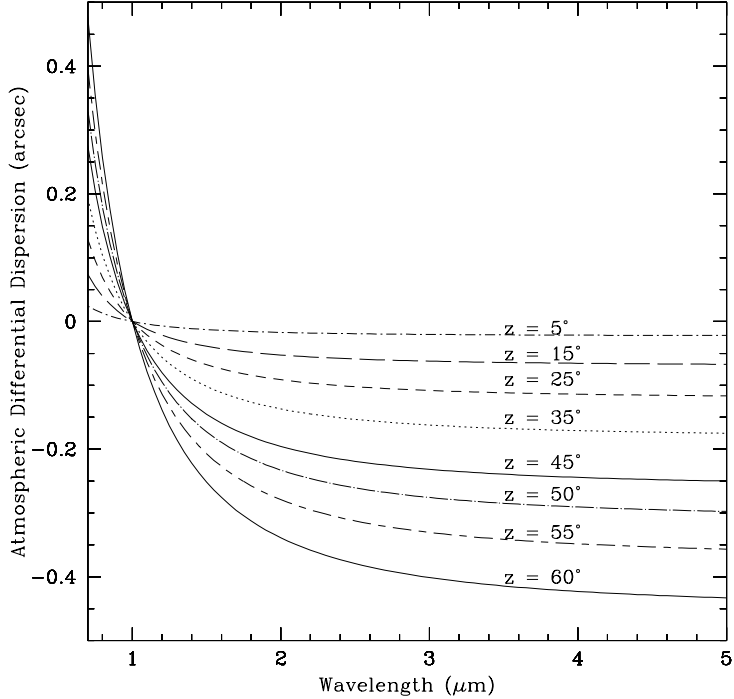}
\caption[]
{Atmospheric differential dispersion vs wavelength at different zenith angle  
above Mt. Hopkins (P = 600  mmHg, T = 280  K, water vapor P = 8 mm Hg).} 	
\end{figure} 

\acknowledgements     

We wish to thank Dr. T. Groesbeck, S. Stahl for useful conversations.  This 
work has been supported by the  AirForce office of Scientific   Research under
grant number F 49620-96-1-0366,  NSF AST-9421311 and AST-9623788.

\newpage
\begin{center}
{\bf REFERENCES}
\end{center} 

\noindent  Beckers, J.M. 1993,  ARA\&A,  31, 13\\
Bacon, R. et al. 1995, A\&AS, 113, 347
Diego, F. et al. 1995, MNRAS, 282, 323\\
Ellerbroek, B.L.  1991, Rep.  RDA-TR23-1502 (R\&D  Associate, Albuquerque, N.M., 1991)\\
Ellerbroek, B.L. 1994, JOSA, 11, 783\\
Ellerbroek, B.L., Pompea, S.M., Robertson, D.J.,  \& Mountain, C.M. 1994, SPIE,
2201, 421\\
Fowler, A.M., et al. 1996, SPIE proceeding, 2816, 150\\
Fried, D.L. 1994, in Adaptive  Optics for  Astronomy (Kluwer  Academic Publishers), eds: Alloin, D.M., \& Mariotti  J.-M.  25\\
Ge, J. et al. 1996, in Adaptive Optics, Vol. 13, OSA  Technical Digest  Series (OSA, Washington  DC), 122\\
Ge, J. et al. 1997, in ESO Workshop on Laser Technology  for Laser Guide Star Adaptive Optics, Carching, in press\\
Greenwood,  D.P. 1979, JOSA,  69, 549\\
Greenwood, D.P. \& Parenti  R.R. 1994, in Adaptive  Optics for  Astronomy (Kluwer  Academic Publishers), eds: Alloin, D.M., \& Mariotti  J.-M. 185\\
Lambert,  D.L. et al. 1990, ApJL, 359, L19\\
Parenti, R.R.   1992, Lincoln Laboratory  Journal,  5, 93
Ridgway, S.T., 1994, in Adaptive  Optics for  Astronomy (Kluwer  Academic Publishers), eds: Alloin, D.M., \& Mariotti  J.-M. 269\\
Rigaut, F. 1994,  in Adaptive  Optics for  Astronomy (Kluwer  Academic Publishers), eds: Alloin, D.M., \& Mariotti  J.-M. 163\\
Sasiela, R.J., Electromagnetic   Wave  Propagation  in Turbulence: A 
Mellin  Transform Approach, (Springer-Verlag, 1993).\\
Sandler, D.G. et al. 1994, JOSA, 11, 925\\
Tull, R.G. 1994, SPIE  proceeding,  2198,  674\\
Vogt, S.S., \& Schroeder, D.J. 1987, in Instrumentation for 
Ground-based Optical Astronomy, Proc.  of  Ninth Santa  Cruz   Summer
Workshop  in Astronomy and Astrophysics,  ed. Robinson, L.B., 3\\
Vogt, S.S., et al. 1994, SPIE  proceeding,  2198, 362\\

\end{document}